\documentclass[review,3p]{elsarticle}

\usepackage{amsmath,amssymb,bm,amsfonts,enumerate,url}
\usepackage{color}
\usepackage{arydshln}
\usepackage{afterpage}
\usepackage{lineno,hyperref}
\usepackage{graphicx}
\usepackage[linesnumbered, ruled]{algorithm2e}\SetKwRepeat{Do}{do}{while}
\usepackage[noend]{algpseudocode}
\usepackage{multirow}
\usepackage{float}
\usepackage{caption}
\usepackage{subcaption}
\usepackage{amsthm}
\usepackage{comment}

\DeclareMathOperator*{\argmax}{\arg\!\max}

\journal{Ad Hoc Networks}

\begin{document}

\begin{frontmatter}

\title{6G-AUTOR: Autonomic CSI-Free Transceiver via Realtime On-Device Signal Analytics\tnoteref{mytitlenote}}
\tnotetext[mytitlenote]{This work was supported by Lockheed Martin Space Systems Company, Lockheed Martin Corporation.}

\author[ncsuaddress]{Shih-Chun Lin\corref{correspondingauthor}}
\ead{slin23@ncsu.edu}
\author[ncsuaddress]{Chia-Hung Lin}
\ead{clin25@ncsu.edu}
\author[ncsuaddress]{K V S Rohit}
\ead{vkanthe@ncsu.edu}
\author[lmcoaddress]{Liang C. Chu}
\ead{liang.c.chu@lmco.com}

\cortext[correspondingauthor]{Corresponding author}

\address[ncsuaddress]{Intelligent Wireless Networking Lab, Department of Electrical and Computer Engineering, North Carolina State University, Raleigh, NC 27695, USA}
\address[lmcoaddress]{Lockheed Martin Space Systems Company, Lockheed Martin Corporation}

\begin{abstract}
Next-generation wireless systems aim at fulfilling diverse application requirements but fundamentally rely on point-to-point transmission qualities. Aligning with recent AI-enabled wireless implementations, this paper introduces autonomic 
radios, 6G-AUTOR, that leverage novel algorithm-hardware separation platforms, softwarization of transmission (TX) and reception (RX) operations, and automatic reconfiguration of RF frontends, to support link performance and resilience. As a comprehensive transceiver solution, our design encompasses several ML-driven models, each enhancing a specific aspect of either TX or RX, leading to robust transceiver operation under tight constraints of future wireless systems. A data-driven radio management module was developed via deep Q-networks to support fast-reconfiguration of TX resource blocks (RB) and proactive multi-agent access. Also, a ResNet-inspired fast-beamforming solution was employed to enable robust communication to multiple receivers over the same RB, which has potential applications in realisation of cell-free infrastructures. As a receiver the system was equipped with a capability of ultra-broadband spectrum recognition. Apart from this, a fundamental tool - automatic modulation classification (AMC) which involves a complex correntropy extraction, followed by a convolutional neural network (CNN)-based classification, and a deep learning-based LDPC decoder were added to improve the reception quality and radio performance. Simulations of individual algorithms demonstrate that under appropriate training, each of the corresponding radio functions have either outperformed or have performed on-par with the benchmark solutions.
\end{abstract}


\end{frontmatter}


\section{Introduction}

6G and beyond will provide global coverage and ubiquitous wireless services~\cite{Akyildiz6g} by meeting strict performance requirements
for diverse industry verticals.
As we move towards the higher generation standards, communication constraints become more stringent. Also, with data-hungry applications breaking the barriers between physical and virtual domains, it is imperative to develop innovative solutions that go beyond merely utilizing data-driven approaches, built for solving specific communication tasks, for achieving superior performance.
Emerged AI/machine learning techniques or data-driven approaches, which can intelligently manage wireless architectures, protocols, and operations through learning from sensory inputs will become a promising enabler in the future. Notably, the European Telecommunications Standards Institute (ETSI) initiated two industry specification groups, experiential networked intelligence, and zero-touch network and service management, working on closed-loop AI mechanisms for network supervisory assistant systems and virtualized network operation automation without human intervention respectively.
O-RAN ALLIANCE~\cite{ORAN} also commits to evolving wireless networks towards more open and smarter deployments by employing similar technologies for autonomous networking and self-management.

Similarly, at PHY and MAC layers, there is a need to introduce Intelligent Radios \cite{IntRad} which enable realtime frontend reconfiguration, and providing protection based on service requirements and environmental conditions. 
Such intelligent autonomic radios employ algorithm-hardware separation to cater to the ever-advancing hardware landscape. 
This is augmented with the data-driven machine learning models that can replicate different transceiver functions, so that pre-trained models replace the complex signal processing involved to realize the same functions.
An advantage of data-aided function replication is that the transmitters and the receivers no longer need to determine the channel state information (CSI) or reference signals (RS) during the whole process. Avoidance of CSI and RS signals provides an added computational benefit by eliminating the excessive overhead, compared to the traditional radio systems, that heavily rely on CSI and RS for successful communication.

\begin{figure}[!t]
    \centering
    \begin{subfigure}[h]{2.6in}
    \centering
        \includegraphics[width=\textwidth]{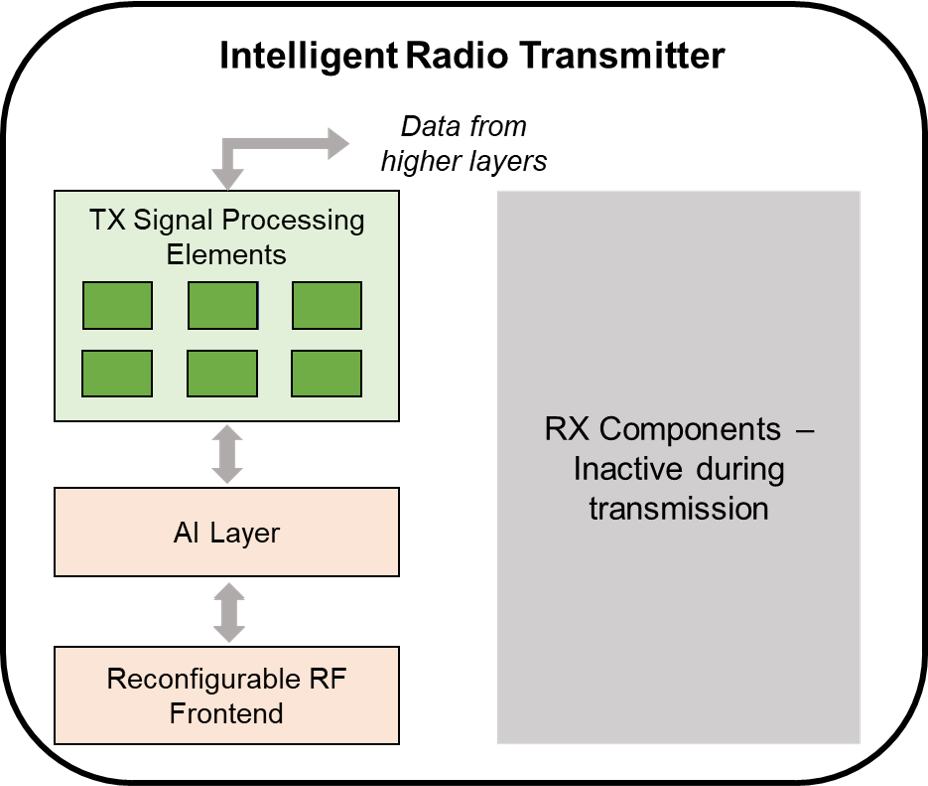}
        \caption{Proposed transceiver as a transmitter}
        \label{tx_block_summary}
    \end{subfigure}
    \hspace{0.5in}
    \begin{subfigure}[h]{2.6in}
        \centering
        \includegraphics[width=\textwidth]{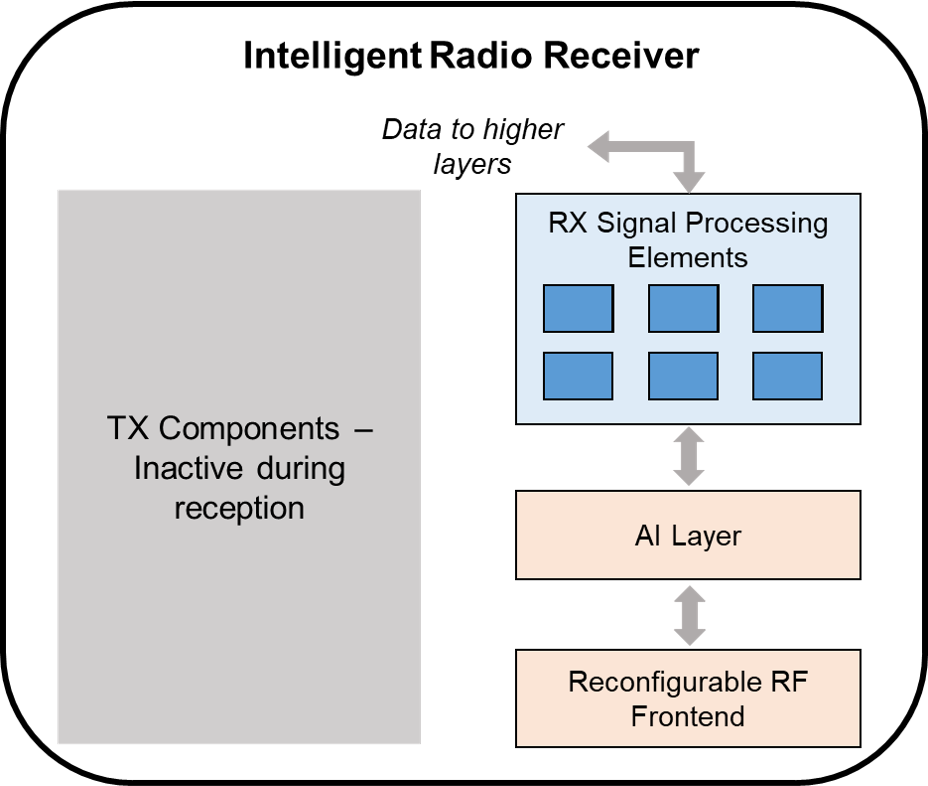}
        \caption{Proposed transceiver as a receiver}
        \label{rx_block_summary}
    \end{subfigure}
    \caption{Intelligent Transceiver with a modular AI layer.}
    \label{block_summary}
\end{figure}

In line with the above requirement, in this paper, we propose an intelligent autonomic radio, 6G-AUTOR, which employs the notion of incorporating algorithm-hardware separation into its design, where algorithms themselves are modelled to be data-driven in nature. We introduce an AI-layer into the transceiver structure, 
which acts as an on-device repository to all machine learning models that are implemented on the radio. Figure \ref{block_summary} shows a high level view of the transceiver working in half-duplex mode, where it will have either of the transmission circuit or the receiver circuit active at a given time. Depending on the mode of operation: transmission or reception, different ML-models will be activated in the AI layer. Specifically, our transceiver utilizes ML-enabled dynamic spectrum access to detect and identify suitable resource blocks during transmission. During reception, it uses a deep learning (DL)-based compression spectrum sensing, feature-based automatic modulation classifier that employs a convolutional neural network (CNN) and yet another DL-based low-density parity-check (LDPC) decoder. Each algorithm that is developed to be a part of the AI-layer has a dedicated function and is independent of other algorithms. For example, an automatic modulation classification block only interacts with the baseband data obtained from the receiver hardware, and informs its classification decision to the demodulation block. And it would not have any influence on other functionalities of the radio, say, spectrum sensing. Similarly, several such disjoint algorithms can be deployed on the common AI-layer, in order to replace complex signal processing functions with ML-based solutions.

\subsection{Algorithm-Hardware Separation}
With the evolution towards 6G, the network deployments will get vastly dense and much more heterogeneous than existing infrastructures \cite{6gUseCases}. Aligning with this development, the underlying hardware cannot afford to remain quasi-static as with current standards \cite{IntRad}. The hardware-software co-design that is in use today, cannot cater to such stringent requirements of the future. In this light, there have been several innovative approaches for increasing the efficiency of the hardware through advancements in MEMS/NEMS technologies \cite{6gHardware_Iannacci}, and by using AI-enabled solutions \cite{6g_hw_ai}.

Algorithm-hardware separation is an attempt to break free from the hardware-software co-design approach that has dictated the telecommunication development since its inception. This paradigm shift enables in building systems that can undergo self-reconfiguration based on the hardware capabilities. This results in the building systems that are robust to changing underlying hardware, by being able to reconfigure the software behaviour accordingly, automatically. Authors in \cite{IoV_Sikdar} have recently proposed such an mechanism to incorporate this novel approach into Internet of Vehicles security frameworks, where the emphasis was on identifying hardware capabilities through iterative testing and machine learning-based decision making process.

Our transceiver is built to support this algorithm-hardware separation, owing to the capability to dynamically reconfigure any data-driven module in real time. Each of the machine learning components, as discussed in subsequent sections, consist of pre-trained neural network models, which are then deployed onto the device for real-time inference. Due to this property, any modification in the algorithms is as simple as updating the model parameters in the AI layer.

\subsection{Relevance to State-Of-The-Art}
\begin{figure}[!t]
\centering
\includegraphics[width=3.2 in]{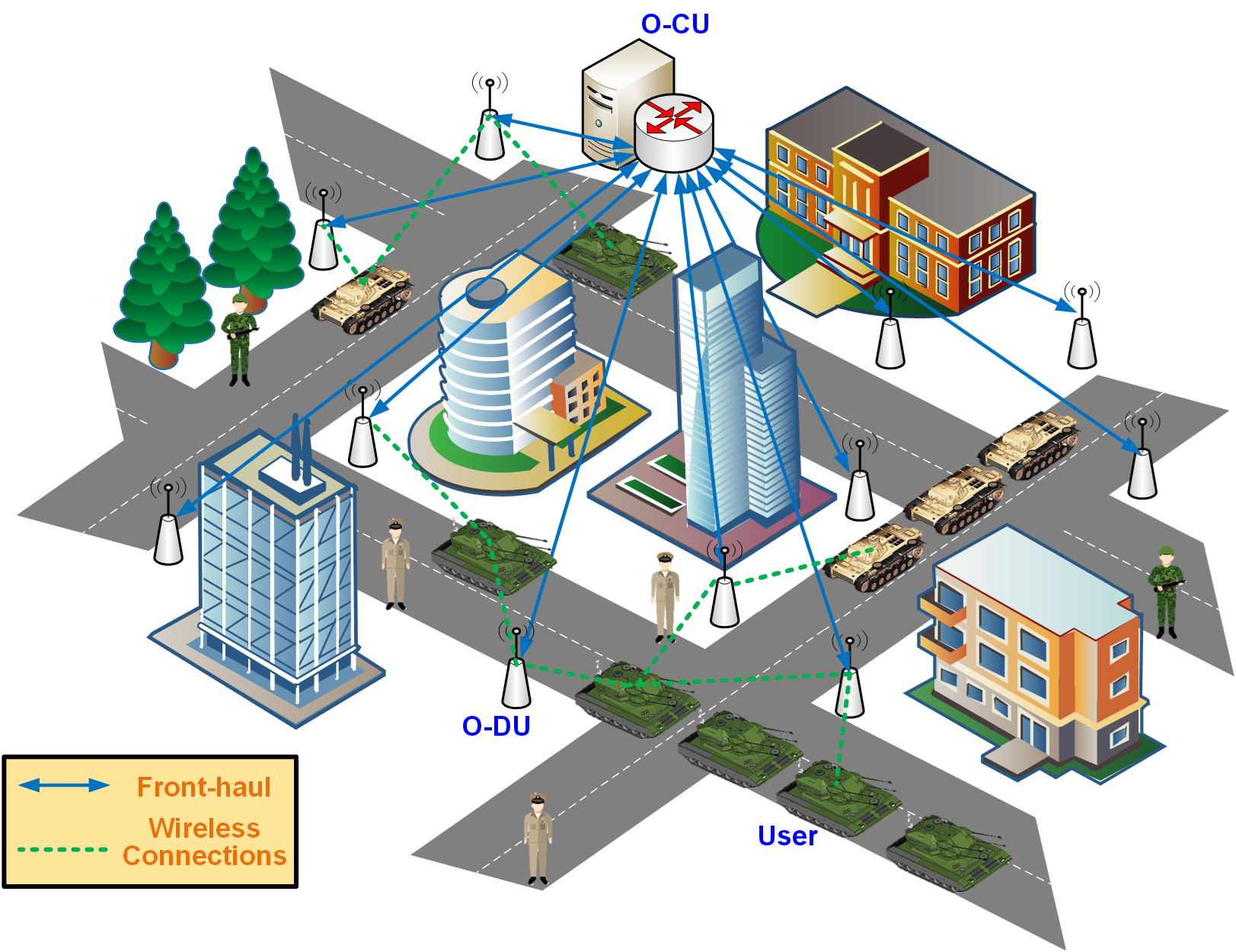}
\caption{The considered cell-free communication system with dynamic spectrum management in Internet of battlefields scenarios. An O-CU controls multiple O-DUs in a geographic coverage area; an O-DU serves many users using the same time-frequency resources for massive machine type communications.}
\label{cell_free_oran}
\end{figure}

In our work, we propose real-time reconfigurability in radios by employing machine learning-based models to replicate RF functions, thereby creating a flexibility of updating/upgrading the models dynamically. This approach aligns with several promising technologies that are well investigated in the development of 5G and beyond systems, such as: dynamic spectrum management, cell-free communications and automatic modulation classification. For example, following the O-RAN structure presented in \cite{ORAN}, in cell-free communications, several Open Distributed Units (O-DUs) (i.e., access points (APs) in 5G systems) controlled by a Open Central Unit (O-CU) (i.e., central processing units (CPUs) in 5G systems) will be deployed to serve users in a geographic coverage area using the same time-frequency resources coordinately as shown in figure \ref{cell_free_oran}, through appropriate dynamic spectrum management. As there are no cells and hence no boundary effects or handover issues, seamless connectivity with reduced transmission overheads can be provided to more users compared to conventional cellular communications. Moreover, to support numerous users simultaneously, static radio spectrum management approach to pre-allocate frequency resources to users is not practical anymore \cite{dsa_textbook}. Alternatively, dynamic spectrum management must be considered to maximize the spectrum utilization rate by exploiting the occasionally idle spectrum. By doing so, it is plausible to support million users with data-hungry applications in a geographic coverage area using the same time-frequency resources. And this can be done computationally efficiently by utilizing the machine learning models to identify suitable resource blocks for transmission and reception. Automatic modulation classification can then be used at the receiver to classify the data into correct modulation type, without any CSI/RS exchange. As a means to increase the end-to-end communication robustness, forward error correction may also be included to compensate for any errors the wireless channel would have caused.

Also, with recent developments in 3GPP Standard Release 17, a new type of communication use case was established: New Radio-Reduced Capacity or NR-RedCap device. This acts as an aggregate of 5G's primary use cases: eMBB, URLLC and mMTC, with certain relaxations such as tolerances to higher latencies, lower throughputs and lower bandwidths \cite{redcap}. 
RedCap devices are aimed towards an implementation primarily in the field of internet of things (IoT), wearables and video surveillance, where the maximum device bandwidth is set at 20 MHz, minimum number of antennas is 1 and latency can extend to as high as 500ms (for video surveillance only).
Our proposed smart transceiver has automation built into it through machine learning model-based inferences, where the models themselves were trained offline. Since neither the model training nor complex signal processing occur on the receiver, our transceivers, even with single-input-single-output (SISO) antenna configuration, can be expected to work conveniently as a RedCap device.

Apart from civilian applications, our system aligns quite well with the tactical communication requirements as well. Considering the Internet of Battlefield Things (IoBT) scenarios \cite{iobt}, the proposed transceiver also can play an essential role to realize command, control, communication, and intelligence systems (C3I) in a real-time manner. First, practical IoBT scenarios may have no ground infrastructure in the field, which makes them suitable for cell-free communications by employing unmanned aerial vehicle as temporal communication infrastructure to serve soldiers and vehicles. Second, military actually owns and shares many spectrum resources with the commercial market. The proposed spectrum recognition module can be used to provide a non-interfering resource block to match the needs for military usages (when used inside a transmitter). Third, the proposed receiver designs can even be used to perform blind eavesdropping attack to obtain information from enemies without any prior knowledge \cite{pradhan}.
Automatic modulation classification is a highly investigated topic for military applications with research in this field aging greater than 2 decades.
Recently, the Department of Defense (DoD) announced \$600 million in awards for B5G experimentation and selected Hill Air Force Base in Utah to realize dynamic spectrum utilization. In March 2022, the DoD has released a Request for Solutions contract that can provide solutions to addess the DoD Tactical Network operations in the presence of 5G network infrastructure \cite{sam_gov}. The DoD has also built prototypes to test out dynamic spectrum management, by the acronym OSCAR (Operational Spectrum Comprehension, Analytics and Response), in 5 aerial combat training ranges, to enable flight emulation training \cite{airforcemag}. These snippets show case the importance of the concepts incorporated in this paper, in terms of their relevance to state-of-the-art implementations in civilian and military applications.

\subsection{System Model}
\begin{figure}[!t]
    \centering
    \begin{subfigure}[h]{6in}
    \centering
        \includegraphics[width=\textwidth]{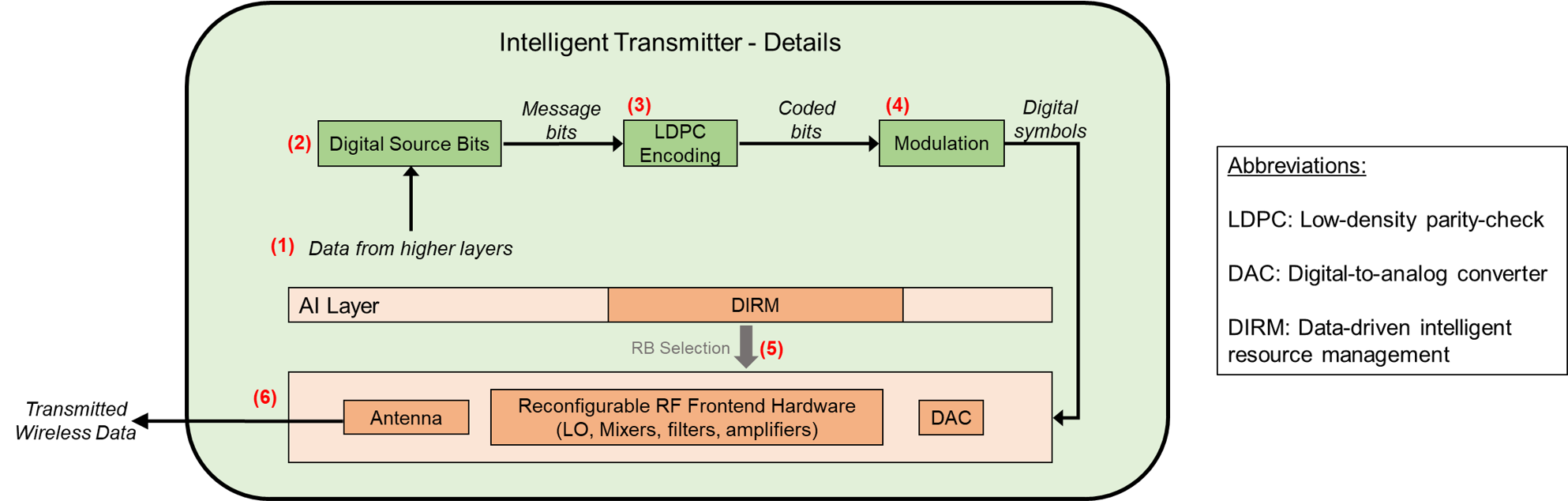}
        \caption{Transmitter with components displayed}
        \label{tx_block_details}
    \end{subfigure}
    \begin{subfigure}[h]{6in}
        \centering
        \includegraphics[width=\textwidth]{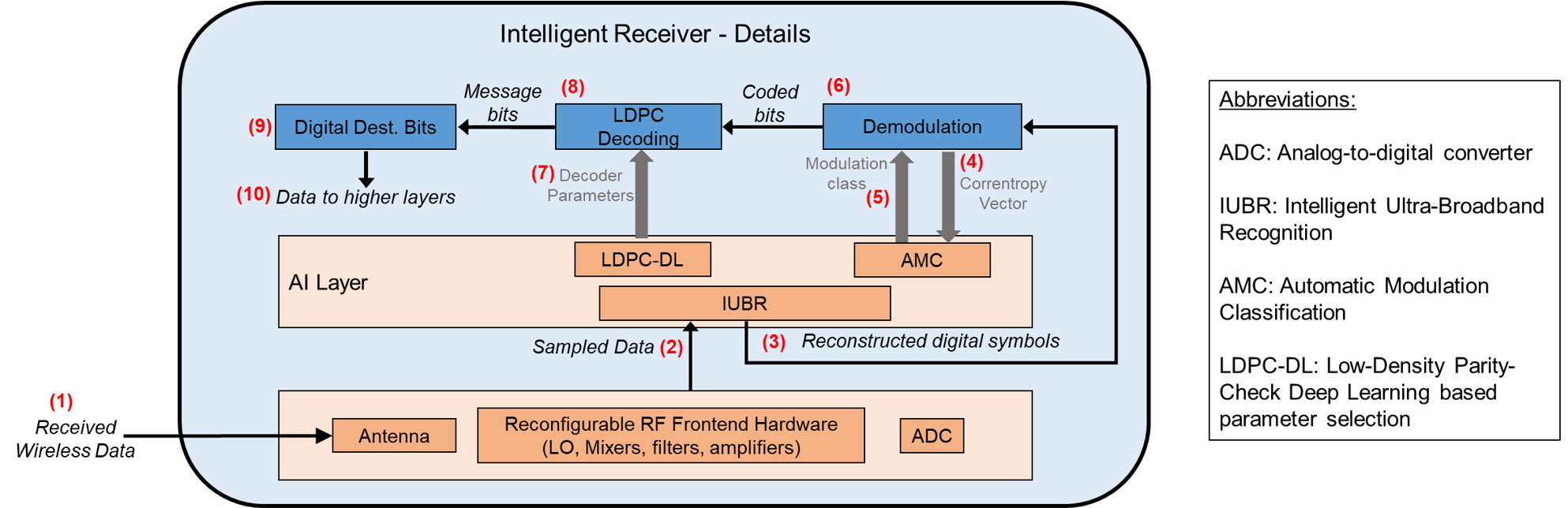}
        \caption{Receiver with components displayed}
        \label{rx_block_details}
    \end{subfigure}
    \caption{Intelligent Transceiver block diagram in detail.}
    \label{block_detail}
\end{figure}

Having introduced the our intelligent autonomic radio structure earlier, this section provides more insights into the specifics of transmitter and receiver functions and the objectives we aim to achieve through such an implementation. Figure \ref{block_detail} provides a detailed description of different components involved in the TX and RX modes of operation shown earlier in figure \ref{block_summary}. As seen here, the AI-layer hosts different models during transmission and during reception. These data-driven models either augment the signal processing tasks by eliminating large computation overheads or provide inferences that eliminate the need for certain modules completely. In this section, we give a high level overview of different data-driven functionalities that our transceiver is capable of, which are detailed in subsequent sections.

When the transceiver acts as a transmitter, it follows the following data-flow: (1) Obtains data from higher layers, (2) process it to create a bit stream,  (3) perform LDPC channel encoding, (4) modulate the coded bits to create symbols, (5) symbols are passed to the RF frontend, where the DIRM (data-driven intelligent resource management) module provides the resource block information for physical transmission, (6) data is sent out using antennas.

When the transceiver acts as a receiver, the following operations occur: (1) data is captured using the antenna, (2,3) it is then sent into the IUBR (intelligent ultra-broadband recognition) module, to perform compressive spectrum sensing and reconstruct the original signal, along with identified carrier frequency, (4) demodulation block performs a complex correntropy feature extraction on the reconstructed data, (5) AMC (automatic modulation classification) module takes the correntropy values and gives out the modulation class as the output, (6) this is followed by a demodulation of based on the classification result, and the data is sent to the LDPC decoder (7) LDPC decoder provides relevant parameters to decoding, (8) decoder performs the iterative decoding using a neural network decoder, and finally (9,10) the resulting message bits are sent to higher layers after appropriate formatting.

It has to be noted that, there are few other components such as the matched filtering, and synchronization blocks were not displayed on the transceiver block diagram, but are an integral part of actual communication realization.

Through this paper, we intend to contribute an intelligent autonomic radio which -
\begin{itemize}
    \item incorporates a dedicated AI layer, which can host multiple machine learning models and can toggle them from active to inactive state, depending on the mode of operation,
    \item showcases the capability to incorporate dynamic spectrum management and cell-free architecture by leveraging the efforts of data-driven, sometimes in conjunction with optimization-driven approaches, to achieve high quality performance,
    \item builds machine learning based-algorithms to augment the complex processing overhead involved in automatic modulation classification and forward error correction using LDPC channel coding and
    \item adheres algorithm-hardware separation principles by enabling the aforementioned data-driven algorithms to be reconfigurable in real-time.
\end{itemize}

The remaining article is structured as follows: Section 2 presents a solution for cross-layer resource management problem to identify suitable resource blocks in multi-agent communication setup (DIRM). Section 3 provides a means to sense the environment using compression spectrum sensing, which can be utilized both on the transmitter and the receiver (IUBR). Section 4 elaborates on the ability to automatically classify the received data into appropriate modulation schemes using feature extraction and CNN-model classification (AMC). Section 5 covers a mechanism to estimate the decoding parameters of the iterative LDPC scheme to recover errors introduced by a wireless channel (LDPC-DL).
Finally, in the last section, we conclude our system design and provide next steps for future work.

\section{Data-driven Intelligent Radio Management
for Proactive Multi-Agent Access}\label{section_2}

We present our transmitter design for distributed cross-layer resource management for 
next generation multi-agent communications.
%
%
%
\subsection{Automatic Reconfiguration and Access}
Resource sensing and selection are two prerequisite steps in conventional multi-agent access, and resource selection can perform purely in a distributed manner based on the sensing results. The current 5G specification still relies on centralized management that collects all sensory information from devices \cite{3gpp38_886}, significantly limiting latency-strict applications. Our objective is to address the cross-layer resource management problem, where users employ distributed resource management without heavy overheads from a centralized infrastructure.

Assuming that there are $k\in\{1,...,K\}$ users forming $K/2$ transceiver pairs in a virtual cell. Based on the half-duplex assumption, we take that, at time slot $t$, half of the users will act as transmitters, and the other users will act as receivers to perform one-side data transmission. At the next time slot $t+1$, the original receiver will become the transmitter and vice versa. At each time slot, all transmitters need to choose a resource block (RB) from $m\in\{1,...,M\}$ RBs in the resource pool and decide on a transmit power to maximize the sum rate of the virtual cell.
To avoid the resource sensing process, each user is expected to finish the above resource management distributively without knowing what the other users are choosing.

Considering user $k'$ that receives data from user $k$ using RB $m$ at time $t$, the signal-to-noise-plus-interference-ratio (SINR) can be expressed as
\begin{eqnarray}
\Gamma_{k'} [m,t] = \frac{P^k[m,t]\times||h_{k'}^k[m,t]||^2}{w[m]\sigma^2+\sum_{i \ne k} P^i[m,t]\times||h_{k'}^i[m,t]||^2},
\end{eqnarray}
where $h[m,t]$ is the channel effect, $w[m]$ is the bandwidth of the RB, and $P^{k}[m,t]$ stands for the transmit power of transmitter $k$.
Following 5G standards, the modulation and coding scheme will be decided according to $\Gamma_{k'} [m,t]$. Hence, the spectrum efficiency $SE[m,t]$ of RB $m$ at time slot $t$ can also be inferred. The user-plane received data bits of this pair can be computed as $r_k[t]=N^{data}_{sym}[m,t] \times SE[m,t]$, where $N^{data}_{sym}[m,t]$ is the number of symbol in RB $m$ at time $t$.

To maximize the sum rate of all transceiver pairs, we formulate multi-agent radio resource management as the following optimization problem: 
\begin{subequations}
\begin{align}
\textbf{\mbox{Find: }}&\ I^k[m,t], P^k[m,t],\ \forall k \label{4a}\\
\textbf{\mbox{Maximize }}&\ \sum_{k} r_k[t] \label{4b} \\
\textbf{\mbox{Subject to }}&\ 0\leq \sum^{M}_{m=1}I^k[m,t]\leq 1,\ \forall k \label{4c} \\
&\ \sum_{m=1}^{M} P^k[m,t]\leq P_{max},\ \forall k \label{4d}
\end{align}
\end{subequations}
Indicator function $I^{k} [m,t]\in\{0,1\}$ describes user $k$'s selection of RB $m$ at time $t$ for communications. Equation \eqref{4c} implies that a transmitter can only select one RB at any given time. Equation \eqref{4d} describes that each user's transmit power cannot exceed a pre-defined power level.

\subsection{The Data-Driven Intelligent Radio Management (DIRM)}\label{performAnal}
We develop a reinforcement learning-based resource management strategy to solve the above optimization. Each user can be regarded as an edge learner (i.e., agent) to learn how to choose RB and transmission power autonomously, based on local observation to maximize the sum rate of the virtual cell.
With the network slicing flexibility provided by software-defined platforms (e.g., \cite{sdvec21}), centralized training can be conducted without any implementation issues. Well-trained weights can then be deployed in each user to perform distributed radio resource management concurrently.

To employ reinforcement learning-based algorithm, we need to define the state, action, and reward function to describe the interested problem as a Markov decision process first. We have the state space with local observation as:
\begin{eqnarray}
    s_t = \{I^{k}_{Tx}[t], g^{k}_{Rx}[t-1], \check{g}^{k}_{Rx}[t-1]\} \ \forall k,
\end{eqnarray}
where $I^{k}_{Tx}[t] \in \{0,1\}$ is an indicator function that defines whether user $k$ acts as a transmitter at time slot $t$, $g^{k}_{Rx}[t-1]$ and $\check{g}^{k}_{Rx}[t-1]$ are the received power and interference power at the RB at time slot $t-1$.

For the action space, each user should make decisions on RB and transmit power selections. Particularly, when a user acts as a receiver, it should choose a dummy RB to receive information. Also, we consider a discrete power-level pool $\mathcal{P} = [-100, 5, 15, 23]$ for transmit power selection to fix action space of each agent with dimension $(M+1)\times|\mathcal{P}|$.
Finally, as for the reward design $r_t$ at time $t$, the sum of spectrum efficiency is set as a reward. The single time slot reward can be extended to the reward of an episode by considering discounted future reward factor $\gamma$, i.e., 
\begin{eqnarray}
    \text{Total Reward} = \sum^{T}_{t=t'}\gamma^{t-t'}r_t.
\end{eqnarray}

To approximate the continuous state and infer the corresponding action to maximize the final reward, various NN-based approaches are available, such as wire-fitter, deep Q-networks (DQN) and the novel double Q-networks in \cite{moon2020}. To avoid overestimating future rewards, we employ the double Q-networks in our algorithm.

A common approach to constructing the edge learner is to use Q-learning.
Each edge learner will have an independent DQN to be trained. Also, a target network (i.e., an accompanying neural network with the same architecture as DQN) will be
used in the training process of all edge learners.
Moreover, we provide $o_t = \{s_t, \epsilon, e\}$ as observations to edge learner to avoid non-stationary problems during training \cite{Replay}. $e$ is the training episode number, and $\epsilon$ is the probability of random action selection.
We obtain the best policy for distributed radio resource management by updating each DQN's weights with
\begin{eqnarray}{rcl}\label{9}
Q^n(o^n_t, a^n_t;\bm{\theta}_n)
\gets  Q^n(o^n_t, a^n_t; \bm{\theta}_n)+\alpha [y_t-Q^n(o^n_t, a^n_t; \bm{\theta}_n)],
\end{eqnarray}
where $y_t = r_t+\gamma  \text{max}_{a'} Q^n(o^n_{t+1}, a'_t; \bm{\theta}_{n'}) $.
In Equation (\ref{9}), each edge learner has a separate DQN $Q^n$ parameterized by $\bm{\theta}_n$ and a target DQN by $\bm{\theta}_{n'}$; both should be trained using this updating equation.
Besides, each agent maintains a memory buffer $\mathcal{D}$ in which they store the tuple $(\{o_{t}^n\},a_t^n,r_t,\{o_{t+1}^n\})$ to reuse the training samples.
Also, the popular $\epsilon$-greedy based action selection with linear annealing mechanism is employed to facilitate the training process.
We summarize the entire DIRM of these centralized training and distributed execution in Algorithm \ref{dirm_algo}.
\begin{algorithm}[htbp]
\SetAlgoLined\SetKwInOut{Input}{Input}\SetKwInOut{Output}{Output}
\Input{Reward $r_i$, action-value functions $\{Q^n(\cdot)\}$, minibatch size $B$, target networks updating frequency $T_{up}$, annealing parameter $\delta$.}
\Output{Access policies $a_{T+1}^n$ for all $1\leq n\leq N$.}
\textbf{compute} annealing rate $\xi=(\tau_{max}-\tau_{min})/\delta T$.\\
\%\% Deep reinforcement learners for proactive access.\\
\textbf{initialize} Q networks $\bm{\theta}_n$ for all $1\leq n\leq N$ randomly. \\
\For{$i=1$ \emph{to} $T$}{
\textbf{compute} $\tau=\max\{\tau_{min}, \xi T\}$.\\
    \For{$n=1$ \emph{to} $N$}{
    \textbf{observe} the environment state $s_i$ for $o_i^n$.\\
    \textbf{select} $a_i^n=1_{\{\tau\leq\bar{\tau}\}}\argmax_{a} Q^n(o_i^n,a;\bm{\theta}_n)+$ $1_{\{\tau > \bar{\tau}\}}\mathcal{U}(\{a_i^n\})$ with uniform function $\mathcal{U}(\cdot)$.\\
    }
\textbf{execute} $a_i=(\{a_i^n\})$ and obtain $r_i$ and $s_{i+1}$.\\
\textbf{observe} state $s_{i+1}$ for $o_{i+1}^n$ for all $1\leq n\leq N$.\\
\textbf{store} $(\{o_{i}^n\},a_i,r_i,\{o_{i+1}^n\})$ in replay buffer $\mathcal{D}$.\\
    \For{$n=1$ \emph{to} $N$}{
    \textbf{sample} a random minibatch of $B$ samples $(\{o_{j}^n\},a_j,r_j,\{o_{j+1}^n\})$ from $\mathcal{D}$.\\
    \textbf{set} $y_j=r_j+\gamma \max_{a'} Q^n(o^n_{j+1},a';\bm{\theta}'_n)$.\\
    \textbf{update} $\bm{\theta}_n$ by minimizing the loss $\mathcal{L}(\bm{\theta}_n)=\frac{1}{B}\sum_j \left(y_j-Q^n(o^n_j,a_j;\bm{\theta}_n)\right)^2$.\\
    \textbf{update} target network $\bm{\theta}'_n$ for every $T_{up}$ as:
    $\bm{\theta}'_n\leftarrow\bm{\theta}_n$.
    }
} 
\caption{Data-Driven Intelligent Radio Management (DIRM).}
\label{dirm_algo}
\end{algorithm}

\subsection{Numerical Results}
We choose vehicular communications (5G NR mode 2)~\cite{3gpp38_886} 
given that vehicular communications are often with more strict latency constraints.
We adopt a well-known Manhattan grid model and use SUMO \cite{SUMO} to generate realistic trajectories of vehicles to evaluate sidelink communications performance.
The SUMO step-size is set as $0.1$ s, the maximum vehicle speed is $45$ miles/hour, and other parameters are listed in Fig.~\ref{5a}.
We use the urban path loss model and shadowing with $5.9$ GHz band in Table $5.2.1$-$1$ in \cite{3gpp38_886}.
We employ DQN hidden layers with the neuron numbers of $400$, $200$, and $150$. 
We consider tanh activation function, RMSProp optimizer with learning rate $1e^{-4}$, buffer size = $25,000$, mini-batch size = $200$, discount factor $\gamma=0.995$, target update counter $\hat{t}_{up}=5$, $\epsilon_{max}=1$, $\epsilon_{min}=0.2$, and $5,000$ training episodes.
The proposed DIRM is compared with the following schemes:
\textbf{Brute Force} (Benchmark): a centralized controller needs to perform an exhaustive search in all possible combinations of the RBs and power allocations to find the best pairs of actions that provide the best sidelink throughput. That is, $\left[\left(M+1\right)\times|\mathcal{P}|\right]^U=(4 \times 4)^6=16,777,216$ possible combinations should be computed in each step.
    While 5G NR mode 2 does not expect any centralized controller, we have the controller in this optimal case to provide the benchmark performance.
\textbf{Random Access} (Naive Baseline): in the random action selection, each agent chooses its action randomly from its possible action space of $O((M+1) \times |\mathcal{P}|)$.

\begin{figure}[!t]
\centering
\subfloat[Vehicular environment setup.]{ 
\footnotesize\begin{tabular}{|c|c|}
\hline
\multirow{2}{*}{Manhattan grid} & 300m$\times$200m   \\
& (with two-way streets)  \\ \hline
Number of vehicles; vehicle speed & 6; 45mph                    \\ \hline
Vehicle antenna height and gain & 1.5m; 3dBi \\ \hline
Vehicle noise figure & 9dB \\ \hline
Maximum vehicle association  & 400m                       \\ \hline
Number of RBs; RB bandwidth & 3; 180kHz            \\ \hline
Transmitted power profile & \{-100, 5, 15, 23\}dBm                   \\ \hline
Noise power & -114dBm/Hz                       \\ \hline
Received SINR margin & (-5, 40)dB                  \\ \hline
Transmission time slot & 1ms               \\ \hline
\end{tabular}
\label{5a}
}\\
\subfloat[Multi-agent radio management.]{ 
\footnotesize\begin{tabular}{|c|c|c|c|}
\hline              & Brute Force   & \multirow{2}{*}{DIRM} & Random      \\
 & (Benchmark) & & Access \\ \hline
Testing duration
& \multicolumn{3}{|c|}{100 episodes = 100s} \\ \hline
Data rates per vehicle pair
& 685Kbps   & 665.5Kbps  & 280Kbps   \\ \hline
Collision rate
& 0\%  & 0.7\% & 31.9\% \\ \hline
Link failure rate
& 0\%  & 0.4\% & 16.6\%   \\ \hline
\end{tabular}
\label{5b}
}
\caption{Average system performance for vehicle PC5 sidelink.}
\label{F1}
\end{figure}
Since the proposed distributed method examines vehicle PC5 sidelink without inheriting any carrier sensing, we show its effectiveness by showing total link collisions and failures.
Fig.~\ref{5b} shows the average collision and link failure rates with $100$ test episodes and fixed SINR threshold $\Gamma_{min}=10$ dB.
\textbf{Brute Force} algorithm provides the performance upper bounds, i.e., the best average data rates and no collisions or link failures during testing via time-demanding resource sensing and consequent centralized resource management. The DIRM's performance is very close to the optimal results no matter which performance indicators. Our distributed algorithm does not access the information obtained from resource sensing, more practical than the benchmark. Also, without sensing-information enhancement, \textbf{Random Access} gives poor performances in all performance metrics, validating the superiority of our DIRM design.

\section{Intelligent Ultra-Broadband Recognition (IUBR)}\label{section_3}

This section describes the proposed intelligent spectrum recognition algorithm, which can be used in both transmitters and receivers to support cell-free communication with dynamic spectrum management.
%
Assuming that $N_s$ subcarriers can be used for wireless data transmission in a geographic coverage area. The goal of spectrum recognition is to identify occupy status of each subcarrier. As a result, 
receivers can 
perform spectrum recognition to identify the subcarriers for following operations, such as automatic modulation classification, to receive information from transmitter successfully.

To realize effective spectrum recognition, a straightforward way is performing Nyquist sampling to check the signal power of each subcarrier. However, due to the wideband nature of B5G communication systems (i.e., sub6GHz, mmWave, and THz bands), performing Nyquist sampling requires high-end ADC and large volume storage to support extremely high sampling rate (may up to several THz) and consequent sampling data, which is impractical for most communication scenarios. Alternatively, compressed sensing technique can aid the on-device spectrum management by lowering the sampling rate significantly. Specifically, the compressed sensing mechanism can be divided into two parts: signal compression and signal reconstruction. In the signal compression stage, original spectrum signal with size as $N_s \times 2$ (i.e., real part and imaginary part) will be condensed as under-sampled measurements with size as $N_m \times 2$ by designing a sensing matrix with the size as $N_s \times N_m$, where $N_m<<N_s$. Then, in the signal reconstruction stage, a signal reconstruction algorithm will be developed to reconstruct the original spectrum signal based on the above under-sampled measurements. By doing so, the requirements of ADC and storage can be alleviated as only under-sampled measurements should be obtained to perform the spectrum recognition.

\subsection{Proposed Algorithm}
Although compressed sensing technique can aid the spectrum recognition process, conventional reconstruction algorithms often contain time-consuming iterations with remarkable computing complexity, being not suitable for the considered on-device scenarios. Alternatively, DL-based reconstruction algorithms are proposed, designing a function to map the under-sampled measurements to the corresponding original spectrum. However, we notice that existing DL-based solutions only focus on the reconstruction part of the compressed sensing framework, employing random projections to generate under-sampled measurements and casing consequent performance bottleneck. 

We present our end-to-end solution~\cite{ISM} to perform joint design of signal compression and signal reconstruction stages, providing superior performance compared to existing DL-based algorithms. 
The compression module contains a specially-designed layer to produce under-sampled measurements by making the trainable weights act as the content of sensing matrix. Then the under-sampled measurements will be fed into the reconstruction module for signal reconstruction. 

There are two features of the proposed algorithm. First, 
one-dimensional fully convolutional neural networks are employed to perform efficient compression and reconstruction. Due to the weight sharing property of convolutional neural networks, the amount of trainable parameters is reduced significantly compare to conventional neural network to perform real-time spectrum analytics. Second, we introduce an end-to-end learning framework to supervise the compression module. This enables designing a structured sensing matrix in contrast to the random projection-sensing matrix employed in existing methods (i.e., unstructured sensing matrix), for achieving better performance. 
Also, using this approach, as the compression module and the reconstruction module are presented as a joint end-to-end learning framework, the functionality of the compression module can be evaluated by the overall reconstruction error (i.e., loss function). By doing so using commonly used algorithms, such as backpropagation, we are able to adjust the trainable parameters in the whole framework (both compression and reconstruction modules), to minimize the reconstruction error, thereby realizing a dedicated design for sensing matrix. Consequently, as more informative under-sampled measurements can be generated by the above design, the neural network architecture can be simplified to enable fast on-device spectrum analytics.

\begin{table}[!t]
    \centering
    \begin{tabular}{c|c|c|c|c}
    \hline
         & \multicolumn{2}{c}{GAN \cite{GAN}} & \multicolumn{2}{c}{Proposed}\\
         \hline
         & sub6 GHz & THz & sub6 Ghz & THz \\
         \hline
         \multicolumn{5}{c}{\textbf{Machine learning performance metrics}} \\
         \hline
         MSE & 0.0685 & 0.0187 & \textbf{0.0026} & \textbf{9.61e-04} \\
         \hline
         Cosine Similarity & 0.3370 & 0.4394 & \textbf{0.9951} & \textbf{0.9908} \\
         \hline
         SSIM & 0.3033 & 0.6289 & \textbf{0.8523} & \textbf{0.9386} \\
         \hline
         \multicolumn{5}{c}{\textbf{Communication performance metrics}} \\
         \hline
         $P_d$ & 2.47e-04 & 0.0622 & \textbf{0.9} & \textbf{0.9449} \\
         \hline
         $P_f$ & 0 & 2.68e-04 & \textbf{2.6e-04} & \textbf{0.0014} \\
         \hline
    \end{tabular}
    \caption{The achieved performance is with SNR = 30 dB and compression rate = 0.125.}
    \label{vcc_}
\end{table}


\subsection{Simulation Results}
Here, we present the performance comparison between the proposed algorithm and state-over-the-art generative adversarial network (GAN)-based spectrum recognition method \cite{GAN}. We choose vehicular environment to test out the spectrum recognition methods. The continuous trajectories of vehicles are generated via SUMO transportation simulation platform \cite{SUMO} and each vehicle will perform vehicle-to-vehicle or vehicle-to-infrastructure communications occasionally by building wireless connections on available subcarriers. Note that the considering simulations are challenging as the number of occupied subcarriers and the power of built connections may vary significantly according to the realistic trajectories generated from SUMO platform. We ran the simulations on both sub6GHz and THz bands to demonstrate the effectiveness of the proposed algorithm. As shown in Fig. \ref{vcc_}, the proposed algorithm outperforms GAN-based algorithm in both machine learning performance and communication performance metrics, irrespective of the chosen frequency band. Note that the proposed algorithm can offer high detection rate and low false alarm rate even in the scenario with high compression rate, making it a strong candidate for our use case.

\section{Automatic Modulation Classification (AMC) with Asynchronous Frontends}\label{section_4}

We propose an on-device AMC algorithm for smart receiver operations, which can detect modulation schemes in practical asynchronours systems.

\subsection{Time, Frequency, and Phase Offsets}
Considering a point-to-point communications system, the received signal $r(t)$ can be expressed as
\begin{eqnarray}{rcl}
r(t) = h(t)\sum^{+\infty}_{j=-\infty}a_jg(t-jT_0-\epsilon T_0)+v(t),
\end{eqnarray}
where $g(t)$ is the raised cosine filter, $T_0$ the symbol period, $h(t)$ the channel effect, $a_j$ the transmitted symbol, $\epsilon T_0\in [0,T_0)$ the time offset at the receiver, and $v(t)\sim S(\alpha,\beta,\gamma,\xi)$ the additive noise following alpha-stable distribution. 
$\alpha$ denotes the characteristic exponent, $\beta$ the symmetry parameter, $\gamma$ the dispersion parameter similar to the variance in Gaussian distribution, and $\xi$ the location parameter similar to the expectation in Gaussian distribution.
Furthermore, the channel effect can be expressed as $h(t)=Ae^{j(wt+\theta)}$, where $A>0$ is the channel amplitude, $\theta \in[-\pi,\pi)$ the channel phase including both the channel phase and residual phase offset, and $w$ the frequency offset at the receiver.
This alpha-stable distribution noise can model natural or artificial impulsive noise, which is ignored in Gaussian noise. The Gaussian noise can be recognized as a particular case of our considered alpha-stable noise.

Our goal is to develop a data-driven algorithm that automatically recognizes correct modulation type of transmitter signals based only on the received asynchronous signal $r(t)$ (having time, frequency and phase offsets) without prior knowledge.
Assume perfect sampling is conducted on the aforementioned received signal. Fig.~\ref{Fig_ConSt} shows the constellations of different signal modulations.

\afterpage{
\begin{figure}[ht]
\centering
\subfloat[BPSK constellation.]
  {\includegraphics[width=1.5in]{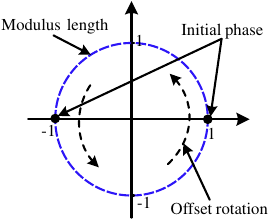}}
\subfloat[QPSK constellation.]
  {\includegraphics[width=1.5in]{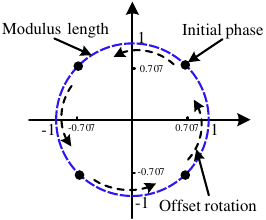}}\\
\subfloat[8PSK constellation.]
  {\includegraphics[width=1.5in]{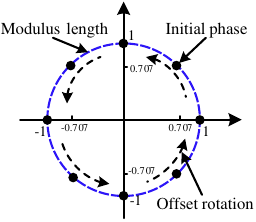}}
\subfloat[16QAM constellation.]
  {\includegraphics[width=1.3in]{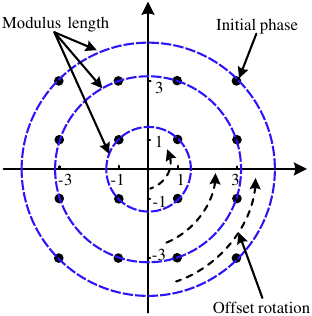}}
\caption{Asynchronous offsets on modulation constellations.}
\label{Fig_ConSt}
\end{figure}
}

\subsection{Low-Complexity On-Device Classifier (LOC)} 
We develop a low-complexity modulation classifier for on-device inference.
Although
conventional deep learning algorithms can directly perform feature extraction, the required computation prohibits on-device AI from executing such procedures. As an alternative, in the developed modulation classifier, we conduct 
feature extraction on the received data, through a cyclic correntropy calculation. This helps in generating discriminating features for each modulation and at the same time, suppress non-Gaussian noise induced from the channel. This dedicated pre-processing design is followed by a convolutional neural network that can classify the discriminating features into appropriate modulation schemes.

For the cyclic correntropy pre-processing stage, a complex Gaussian kernel $\kappa_\sigma$
is defined such that,
\begin{equation*}\label{6}
  \kappa_\sigma(r(t)-r(t+\tau)) = \frac{1}{\sqrt{2 \pi} \sigma^2} \mathrm e^{-\frac{1}{2\sigma^2}\left(r(t)-r(t+\tau)\right)\left(r(t)-r(t+\tau)\right)^*}.
\end{equation*}
We then define the complex correntropy function $V_r(t, \tau)$ as
\begin{eqnarray}
 V_r(t,\tau) &=& \mathbf{E} \Big [ \kappa_\sigma\big(r(t)-r(t+\tau)\big )  \Big ], \\
             &=& \mathbf{E} \Big [ \frac{1}{\sqrt{2 \pi} \sigma^2}\mathrm \mathrm e ^{-\frac{1}{2\sigma^2}(r(t)-r(t+\tau))(r(t)-r(t+\tau))^*}  \Big ].
\end{eqnarray}
where $\mathbf{E}[\cdot]$ is the expectation operator, $\sigma$ denotes complex Gaussian kernel size.
To deal with discrete-time constellations after sampling, we can compute the discrete-version complex correntropy function by
\begin{eqnarray}
V[m] 
     = \frac{\sum_{n=m}^{N} \mathrm e ^{-\frac{1}{2\sigma^2}(x[n]-x[n-m])(x[n]-x[n-m])^* }}{(N-m+1)\sqrt{2 \pi} \sigma^2}.
\end{eqnarray}

This completes the preprocessing and feature extraction, and the results will be used for final modulation classification.

Since the feature sets obtained above are one dimensional sequences, a 1D-convolutional neural network is a good fit for such a dataset. We employed a Conv1D model from the TensorFlow library \cite{keras_conv} to create a 1D-CNN model and train it using synthesized data. The network consisted of two convolution layers, followed by an average pooling layer to decrease the dimension of the features, and ended with a dense layer which invoked a softmax operator on its input. The softmax operation results in an estimation of the probability of the message belonging to a particular modulation class. For an input message $y$, if $\hat{y}$ is the predicted output, then the the loss function $E$ was calculated and network parameters were updated using stochastic gradient descent and backpropagation.
\begin{align*}
    E = -\sum_n \sum_m y_{nm} \log \hat{y}_{nm},
\end{align*}
where $m$ is a numerical (integer) representation of modulation class, and $n$ represents the $n^{th}$ the training set.

We summarize the whole LOC training process in Algorithm \ref{amc_algo}.
\begin{algorithm}[htbp]
\SetAlgoLined\SetKwInOut{Input}{Input}\SetKwInOut{Output}{Output}
\Input{Received I/Q data, i.e., $\bm{I}=\left\{{I}_1,{I}_2,\cdots,{I}_m \right\}$ and $\bm{Q}=\left\{{Q}_1,{Q}_2,\cdots,{Q}_m \right\}$, $m=10000$.}
\Output{Signal modulation scheme $\{\hat{z}_i;1\leq i\leq m\}$.}
\%\% Complex correntropy feature extraction.\\
\textbf{set} kernel $\sigma$ and length $l$ of complex correntropy features for modulation schemes (e.g., $l=2000$).\\
\For{$i=1$ \emph{to} $l$}{
      $\bm{Y}_1=\bm{Y}(1:m-i+1)=\{y_1, y_2, \cdots, y_{m-i+1} \}$ and $\bm{Y}_2=\bm{Y}(i:m)=\{y_i, y_{i+1}, \cdots, y_{m} \}$, where $\bm{Y}=\bm{I}+j\bm{Q}=\left\{y_1,y_2,\cdots,y_m \right\}$. \\
      \textbf{compute} complex correntropy $V[i] = \frac{1}{m-i+1}\sum_{n=1}^{m-i+1} \kappa_\sigma(\bm{Y}_1[n]-\bm{Y}_2[n])$,
     where $\kappa_\sigma(\cdot)$ is a Gaussian kernel.
}
\%\% Identify the modulation scheme via 1D-CNN.\\
\textbf{initialize} weights $\bm{W}^{(0)}$ and biases $\bm{b}^{(0)}$ randomly. \\
\For{$i=1$ \emph{to} $T$}{
\textbf{compute} the first convolutional layer with input $\bm{V}$ and output $\bm{C}_1$, and the second convolutional layer with output $\bm{C}_2$ and kernel size $100$ via the forward-propagation method.\\
\textbf{compute} the pooling layer with output $\bm{C}_p$ by averaging $\bm{C}_2$, and the dense layer with output $^{K\backslash 1}\bm{C}_d$ and total $K$ of modulation schemes.\\
\textbf{get} predicted output $\hat{z}_i=\argmax_{1\leq k\leq K}P_k$ for all $1\leq i\leq m$ via a softmax classifier with normalized likelihood $P_k=e^{\bm{C}_d[k]}/\sum_j e^{\bm{C}_d[j]}$.\\
\textbf{compute} the predicted error from cross-entropy loss $L(\bm{W}^{(i)},\bm{b}^{(i)})$.\\
\textbf{update} 1D-CNN using stochastic gradient decent and learning rate $\eta$ via the back-propagation:
$\left\{\begin{array}{l}
{{\bm{W}}^{(i+1)}} = {{\bm{W}}^{(i)}} - \eta \frac{\partial }{{\partial {{\bm{W}}}}}L\left( {{\bm{W}}^{(i)},{\bm{b}}^{(i)}} \right)\\
{{\bm{b}}^{(i+1)}} = {{\bm{b}}^{(i)}} - \eta \frac{\partial }{{\partial {{\bm{b}}}}}L\left( {{\bm{W}}^{(i)},{\bm{b}}^{(i)}} \right)
\end{array}\right..$
}
\caption{Low-Complexity On-Device Classifier (LOC).}
\label{amc_algo}
\end{algorithm}

\subsection{Numerical Results}
We performed numerical verification of the algorithm in two stages: simulation and over-the-air (OTA) communication. In simulation, a training dataset was created using modulation scheme on a PC for BPSK, QPSK and 16-QAM, with AWGN channel-based propagation at different SNR values: $0,2,4,6,8,10$ dB and possessing multipath: $1,2,4$ paths. $50$ messages were generated at random for each combination of modulation and SNR values, to form the complete training dataset. Correntropy sequences of size $150$ taps were extracted from each modulated message, which passed as training data to 1D-CNN.

After successful offline training, we pass randomly generated messages through a GNURadio-simulated AWGN channel, at different noise levels: $0, 100, 200, 300, 400, 500$ (mV) and at varying normalized frequency offsets: $0, 0.005, 0.01, 0.03, 0.05$ (with respect to the central frequency - 915 MHz). The resulting data obtained at the receivers were passed through the feature extraction and a trained 1D-CNN to determine the modulation classification.

\begin{figure}[!t]
    \centering
    \includegraphics[width=3.2in]{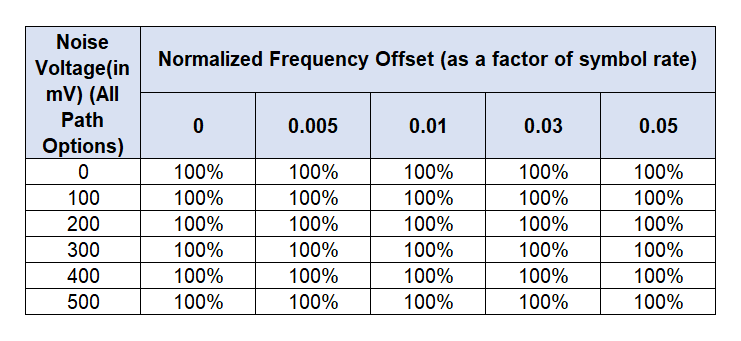}
    \caption{AMC classification accuracy in simulation}
    \label{bmc_sim_result}
\end{figure}

Figure \ref{bmc_sim_result} shows the classification accuracy of the model for all three modulation schemes BPSK, QPSK and 16-QAM.

\begin{figure}[!t]
    \centering
    \begin{subfigure}[h]{3.2in}
    \centering
        \includegraphics[width=\textwidth]{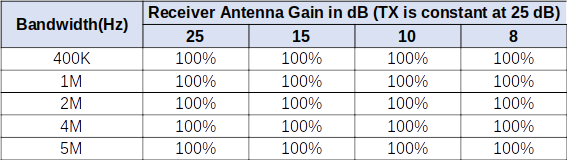}
        \caption{AMC accuracy for BPSK for OTA communication}
        \label{bpsk_bmc_ota_result}
    \end{subfigure}
    \begin{subfigure}[h]{3.2in}
        \centering
        \includegraphics[width=\textwidth]{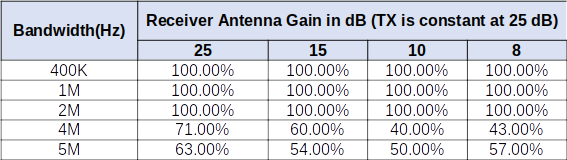}
        \caption{AMC accuracy for QPSK for OTA communication}
        \label{qpsk_bmc_ota_result}
    \end{subfigure}
    \label{bmc_ota}
    \caption{AMC Performance}
\end{figure}

In the second stage, wireless communication data was collected through over-the-air measurements using two USRP 2901s placed in close proximity to each other. Now, the random generated data was passed over a true wireless channel,through SDRs tuned to transmit and receive at a central frequency of 915 MHz, with varying bandwidths: $0.4, 1, 2, 4, 5$ MHz and varying RX antenna gain: $25, 15, 10, 8$ dB. Transmit antenna gain was maintained at a constant value of $25$ dB. Figures \ref{bpsk_bmc_ota_result} and \ref{qpsk_bmc_ota_result} show the performance of AMC on OTA data.

As the bandwidth increased, the correntropy features for BPSK and QPSK seemed very similar, which is why the 1D-CNN model was unable to provide a 100\% accuracy in QPSK modulation scheme. We hypothesize that this similarity could be due to increased noise floor for higher bandwidths, as the classification through correntropy calculation has an inherent assumption that the noise is not too large compared to the actual signal. At higher bandwidths, since this constraint cannot be guaranteed, the features may sometimes seem indistinguishable for different modulation schemes.

\section{LDPC Decoder Parameter Estimation using Deep Learning}\label{section_5}

Low-Density Parity-Check codes have outstanding performance characteristics in terms of block error rates against the channel SNR. While these codes have been around for a long time, their adoption into the 5G standards have resulted in development of techniques that can mitigate the high computational complexity of the algorithm using deep-learning methods \cite{ldpcMain} and by modifying the decoding architectures to enable implementation on devices such as FPGAs \cite{nadal2021}. Conventional method in LDPC decoding include the iterative message passing algorithms such as sum-product (SP) or min-sum (MS) approach. There have been a few efforts in the direction of completely eliminating the iterative aspect by considering the coded message at the receiver as a text and using convolutional neural networks for mapping the message to the correct transmitted code \cite{yanqin2019}. Another approach adopted is to represent the decoding iterations in terms of neural network layers and improve the performance of this neural network using parameter tuning. This process involves attaching weights and biases to either the messages being passed or as normalization and offset components \cite{ldpcMain,qing2020}.

LDPC codes are a form of soft-in soft-out linear codes, where the operation adds a certain number of redundant bits to a binary message of size ‘k’ bits, resulting in a ‘N ’ bit codeword ($N > k$). This increases the robustness to errors due to harsh channel environments.

\subsection{Neural MS Decoder}
In this paper, we have adopted the neural decoder model presented in \cite{ldpcMain}, which has the following characteristics:
\begin{itemize}
    \item A neural network with $N$ input nodes, representing $N$ bits of encoded codeword.
    \item Each iteration in a conventional tanner-graph based decoder is unfolded into a hidden layer.
    \item Each hidden layer with size equal to number of edges in the protograph-LDPC parity check matrix, as per the 5G standard \cite{3gpp212}.
    \item Within each hidden layer, a variable node and check node update are implemented.
    \item Weights and biases of every layer represent the normalization and offset parameters of the decoder.
\end{itemize}

Consider an edge in the parity check matrix $e = (v,c)$, where $v$ represents a variable node and $c$ represents a check node. Then if $\mathcal{E} = \{e=(v,c)\}$ is a set of all edges, and if the LLR inputs into the neural network are given as $l_v$ since each variable node corresponds to a code bit, then for $i^{th}$ hidden layer:
\begin{eqnarray}
    \text{Variable node update for edge $e$: } l_{e=(v,c)}^{i_v} = l_v + \sum_{e' = (v,c'),c' \ne c} l_{e'}^{(i-1)_c},
\end{eqnarray}
\begin{eqnarray}
    \text{Check node update for edge $e$: } l_{e=(v,c)}^{i_c} =& \Bigg(\prod_{e' = (v,c'),v' \ne v} \mathrm{sgn}\left(l_{e'}^{i_v}\right) \Bigg) \\ \notag
    &\times \mathrm{ReLU}\Big( \alpha_{e}^{i} \times \min_{e' = (v,c'),v' \ne v} \big|l_{e'}^{i_v}\big| - \beta_{e}^{i} \Big),
\end{eqnarray}
where $\alpha^{i}$, the normalization factor, is represented by weights of $i^{th}$ hidden layer and $\beta^{i}$, the offset factor, is represented by bias of $i^{th}$ hidden layer.

For reducing the number of computations, we employed a parameter-sharing strategy where one value of $\alpha_e^i$ and $\beta_e^i$ is used within one hidden layer. That is, $\forall \, e \in \, \mathcal{E}$, we have $\alpha_e^i = \alpha^i \, , \, \beta_e^i = \beta^i$. This helps in expanding the network for especially large message sizes, which would need many hidden layers for effective decoding, without linearly increasing the parameter count, thereby restricting the computational complexity as the message size scales up.

\begin{figure}[!t]
    \centering
        \includegraphics[width=6in]{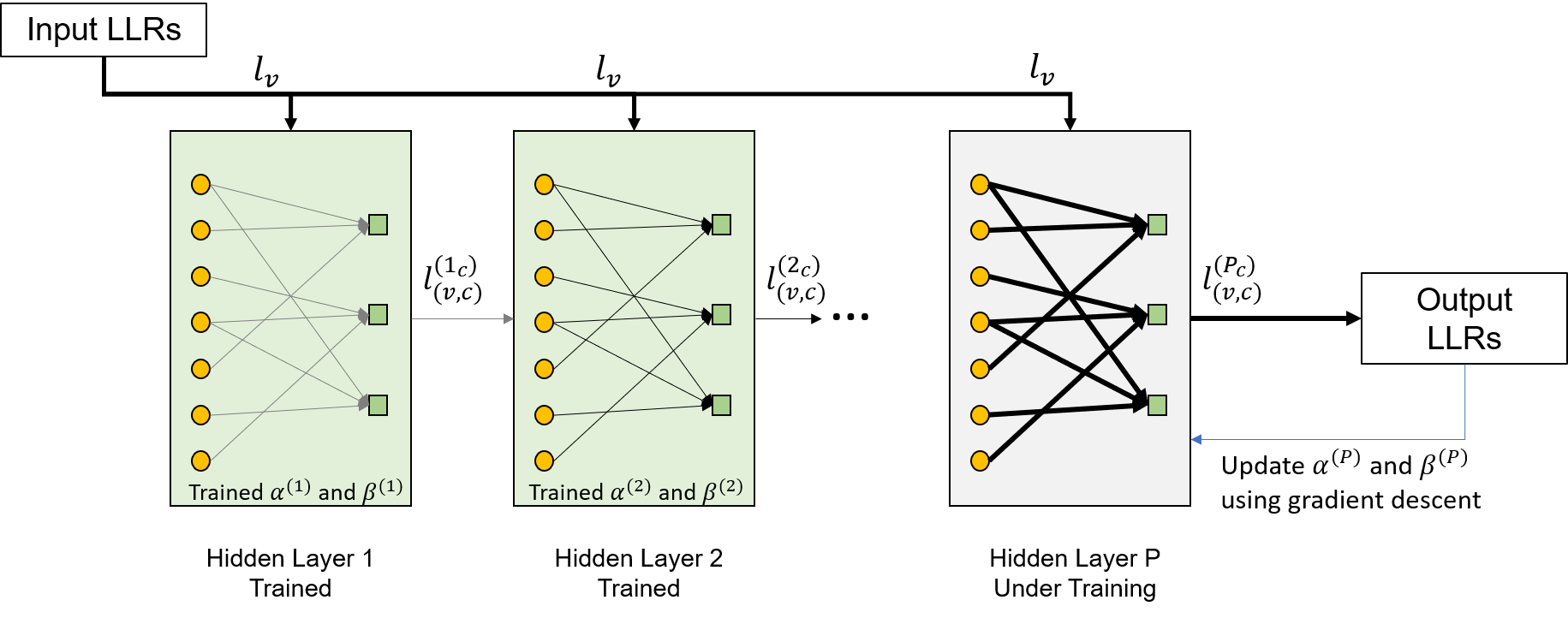}
    \caption{Iteration-by-iteration training}
    \label{hidden_layer_training}
\end{figure}

\subsection{Training the Neural MS Decoder}
For the purposes of training, Basegraph BG2, defined in \cite{3gpp212} was chosen as the parity check matrix for implementing the decoder. A lifting factor of $Z = 16$ was used over BG2. This results in a codeword of size $52 \times 16 = 832$ bits. Random messages of size $832$ bits were synthesized to use in training. A network of $25$ hidden layers was created and trained over the synthesized codewords. Cross-entropy was used a loss function to train the neural network.
\begin{eqnarray}
    L = -\frac{1}{N} \sum_{v=1}^{N} x_v \log(o_v) + (1 - x_v) \log(1 - o_v),
\end{eqnarray}
where $N$ is the number of bits, $o_v$ is output from the network and $x_v$ is the training label used to obtain $o_v$.

In order to eliminate the problem of vanishing gradients in deep neural networks, an iteration-by-iteration training was applied. In this operation, parameters in each layer are trained independently using gradient descent and a cross-entropy loss. After the first hidden layer is trained and its parameters are optimized, second hidden layer would undergo training. However, the computation of LLR outputs during training will now include both the trained first layer and the second layer. Same principle is propagated over all the layers. This implies that the network parameters are optimized by leveraging multiple layers at once, while training only one layer at a time. This results in reduced complexity in computation and in eliminating vanishing gradient problem. Figure \ref{hidden_layer_training} shows the data flow when training the $P^{th}$ hidden layer using iteration-by-iteration method.

\begin{figure}[!t]
    \centering
    \includegraphics[width=0.95\textwidth]{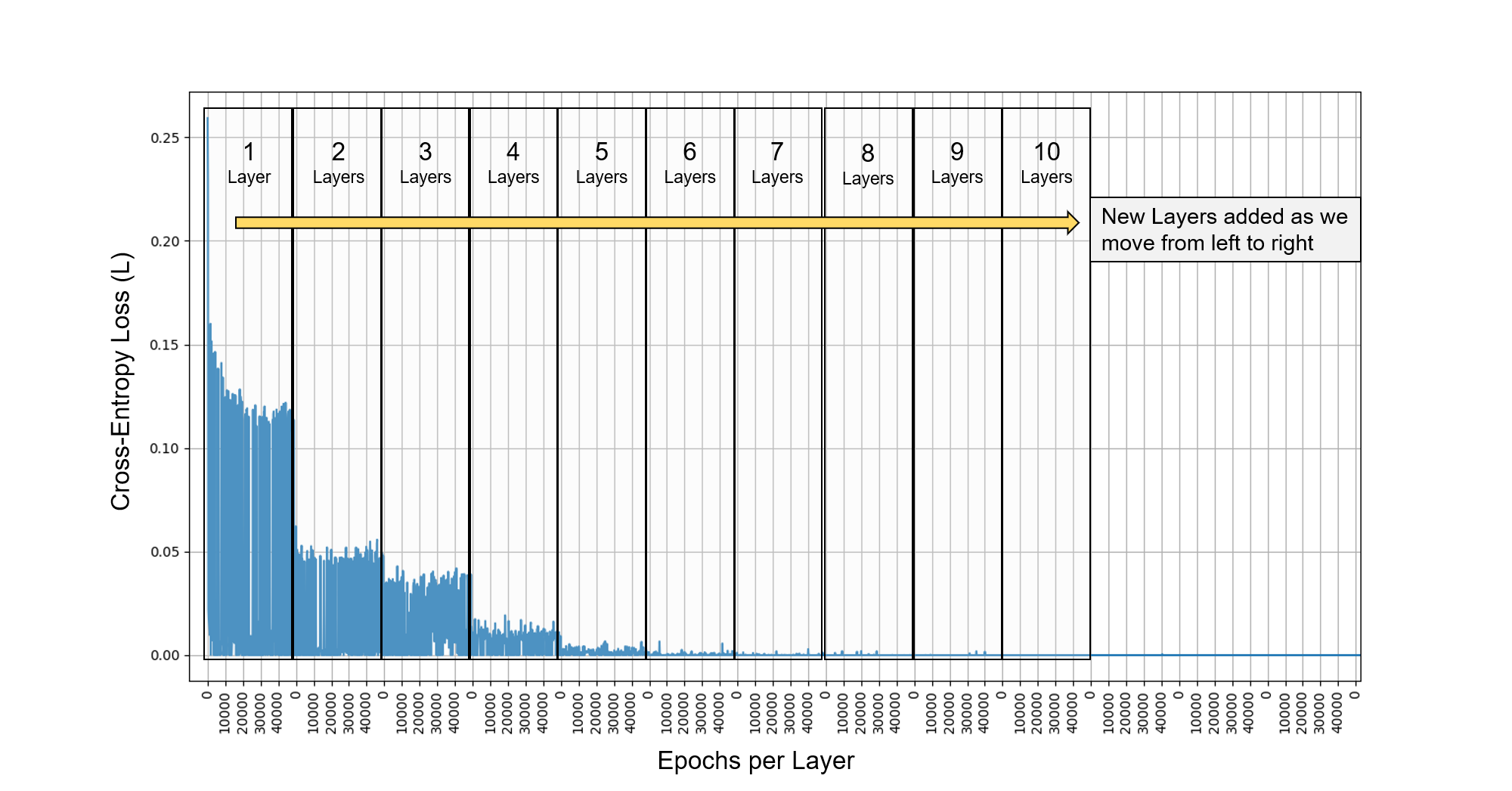}
    \caption{Loss function vs hidden layers (Iterations)}
    \label{neural_ms_training_loss}
\end{figure}

Figure \ref{neural_ms_training_loss} shows the gradual reduction in loss values as the number of layers are increased. While the network was originally built with $25$ hidden layers, for a codeword of $832$ bits, the network converged for less than $10$ layers. A trade-off between computation time and accuracy can be deduced from the loss values over training epochs and multiple layers. Using a network with higher number of layers results in a proporational increase in computation time, due to having to perform the iterative decoding process that many times. However, higher the number of layers used, greater is the accuracy of prediction, since the corresponding cross-entropy loss tends to be lower. Depending on the capability of the device under consideration, one may choose to exploit this trade-off to implement efficient and timely decoding. Since the decoding operations are performed over a PC in our project, we chose to utilize the complete set of $25$ layers in our inference model. This layer count can be decreased to a smaller number when the algorithm has to be ported onto low-complexity devices such as SDRs. This flexibility, enabled due to iteration-by-iteration training, makes this model relevant to algorithm-hardware design and for deploying in RedCap devices.

\subsection{Numerical Results}
\begin{figure}[!t]
    \centering
    \begin{subfigure}[h]{6.4in}
    \centering
        \includegraphics[width=\textwidth]{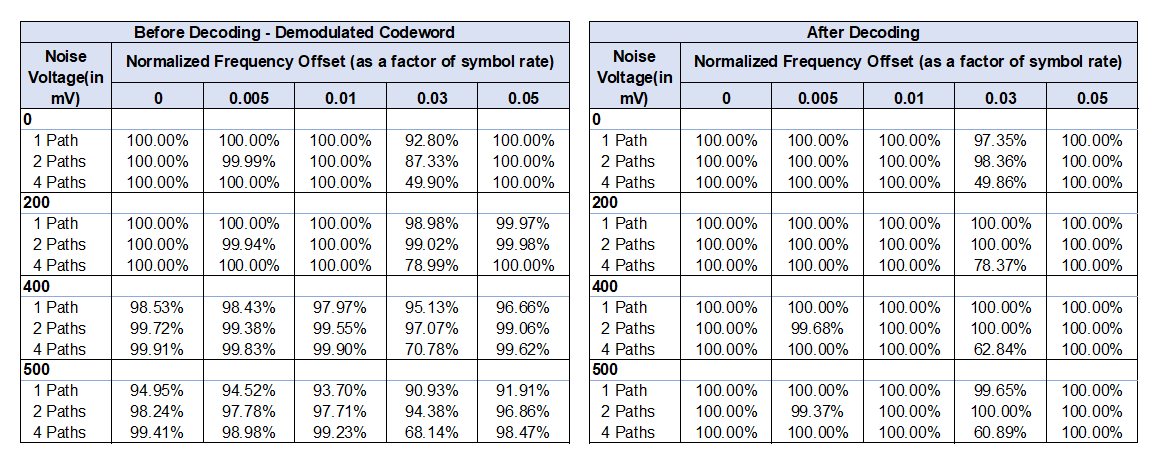}
        \caption{\% of correct bits before and after LDPC decoding in simulation setup}
        \label{qpsk_ldpc_sim}
    \end{subfigure}
    \begin{subfigure}[h]{3.2in}
        \centering
        \includegraphics[width=\textwidth]{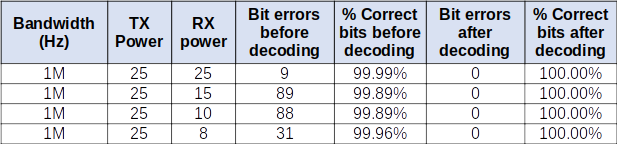}
        \caption{LDPC decoder performance in OTA setup - 1 MHz bandwidth}
        \label{qpsk_ldpc_ota}
    \end{subfigure}
    \caption{LDPC Decoder Performance}
\end{figure}

Similar to the evaluation metrix used for AMC, verification of LDPC decoding was executed in two stages: simulation and OTA communication. Here, two modulation schemes were considered: BPSK and QPSK, transmitted over a GNURadio-based AWGN channel with varying noise levels: $0,200,400,500$ mV, different frequency offsets: $0, 0.005, 0.01, 0.03, 0.05$ (with respect to central frequency of 915 MHz) and multiple paths: $1,2,4$ paths. For OTA communication, BPSK and QPSK modulated data was considered, passing through a wireless channel with bandwidths: $0.4,1,2$ MHz with TX antenna gain at $25$ dB and RX antenna gain varying as $25, 15, 10, 8$ dB. Figures \ref{qpsk_ldpc_sim} and \ref{qpsk_ldpc_ota} show that the LDPC decoder has suppressed most of the errors that were induced by the channel, in simulation and over the air respectively.

\section{Conclusion and Future Work}
This paper introduced intelligent autonomic radios with architecture that enables algorithm-hardware separation, while providing a dedicated and scalable AI layer to accommodate data-driven machine learning solutions to augment the classical signal processing blocks in PHY and MAC layers of the radio. We have also provided a detailed use case involving the introduced components, by fitting them into a transmitter and a receiver data flow, for providing a means of employing the proposed transceiver in an end-to-end communication setup. As a future work, we are working towards improving the AMC algorithm performance at large bandwidths. It is important to maintain the complexity of our algorithm as hardware-friendly as possible, while mitigating the correntropy assumption limitations at large bandwidths. We are also considering an end-to-end system realization over software defined radios, as a hardware proof of concept of the introduced intelligent autonomic radio design. Apart from algorithm/hardware enhancement, another potential future direction is to transform the current algorithms to be capable of self-reconfiguration, based on the underlying hardware utilized. This is expected to be accomplished by incorporating an operating system (OS) that can mediate hardware status in real time to the algorithms. In such a scenario, algorithms would be built as software over the OS. Alternatively, we are keen to explore the possibility of building containers over the device representing the algorithms, whose resources can be modified based on the hardware capability at the time. This would result in building a platform that would cater to 6G requirements as Functions-as-a-service block.

\bibliography{mybibfile}

\end{document}